\documentclass[%
 reprint,
superscriptaddress,
 amsmath,amssymb,
 aps,
 prl,
]{revtex4-2}

\usepackage{graphicx}% Include figure files
\usepackage{dcolumn}% Align table columns on decimal point
\usepackage{bm}% bold math
\usepackage{xcolor}
\usepackage{amssymb}
\usepackage{wasysym}
\usepackage{lineno}
\usepackage{appendix}

\definecolor{sperm1}{rgb}{0, 0, 0.714285714285714}
\definecolor{sperm2}{rgb}{0, 0, 0.857142857142857}
\definecolor{sperm3}{rgb}{0, 0, 1}
\definecolor{sperm4}{rgb}{0, 0.142857142857143, 1}
\definecolor{sperm5}{rgb}{0, 0.285714285714286, 1}
\definecolor{sperm6}{rgb}{0, 0.428571428571429, 1}
\definecolor{sperm7}{rgb}{0, 0.571428571428571, 1}
\definecolor{sperm8}{rgb}{0, 0.714285714285714, 1}
\definecolor{sperm9}{rgb}{0, 0.857142857142857, 1}
\definecolor{sperm10}{rgb}{0, 1, 1}
\definecolor{sperm11}{rgb}{0.142857142857143, 1, 0.857142857142857}
\definecolor{sperm12}{rgb}{0.285714285714286, 1, 0.714285714285714}
\definecolor{sperm13}{rgb}{0.428571428571429, 1, 0.571428571428571}
\definecolor{sperm14}{rgb}{0.571428571428571, 1, 0.428571428571429}
\definecolor{sperm15}{rgb}{0.714285714285714, 1, 0.285714285714286}
\definecolor{sperm16}{rgb}{0.857142857142857, 1, 0.142857142857143}
\definecolor{sperm17}{rgb}{1, 1, 0}
\definecolor{sperm18}{rgb}{1, 0.857142857142857, 0}
\definecolor{sperm19}{rgb}{1, 0.714285714285714, 0}
\definecolor{sperm20}{rgb}{1, 0.571428571428571, 0}
\definecolor{sperm21}{rgb}{1, 0.428571428571429, 0}
\definecolor{sperm22}{rgb}{1, 0.285714285714286, 0}
\definecolor{sperm23}{rgb}{1, 0.142857142857143, 0}
\definecolor{sperm24}{rgb}{1, 0, 0}
\definecolor{sperm25}{rgb}{0.857142857142857, 0, 0}

\begin{document}

\preprint{APS/123-QED}

\title{Dispersion Relations for Active Undulators in Overdamped Environments}

\author{Christopher J. Pierce}
\affiliation{School of Physics, Georgia Institute of Technology}
\affiliation{School of Chemical and Biomolecular Engineering, Georgia Institute of Technology}
\author{Daniel Irvine}
\affiliation{Department of Mathematics, Kennesaw State University}
\author{Lucinda Peng}
\affiliation{School of Chemical and Biomolecular Engineering, Georgia Institute of Technology}
\author{Xuefei Lu}
\affiliation{School of Chemical and Biomolecular Engineering, Georgia Institute of Technology}
\author{Hang Lu}
\affiliation{School of Chemical and Biomolecular Engineering, Georgia Institute of Technology}
\author{Daniel I. Goldman}
\affiliation{School of Physics, Georgia Institute of Technology}

\date{\today}

\begin{abstract}
Organisms that locomote by propagating approximately sinusoidal waves of body bending maintain performance across different environmental substrates by modifying the frequency $\omega$ or wavenumber $k$ of their gait. We identify a unifying relationship between these parameters for overdamped undulatory swimmers (including nematodes, spermatozoa, and mm-scale fish) moving in diverse environmental rheologies, in the form of an active `dispersion relation' $\omega\propto k^{\pm2}$. A model treating the organisms as actively driven viscoelastic beams reproduces the experimentally observed scaling. The relative strength of rate-dependent dissipation in the body and the environment determines whether $k^2$ or $k^{-2}$ scaling is observed. The existence of these scaling regimes reflects the $k$ and $\omega$ dependence of the various underlying force terms and how their relative importance changes with the external environment and the neuronally commanded gait.

\end{abstract}

\maketitle

Organismal self-propulsion results from the cyclical self-deformation of a body in space and time. In overdamped mechanical regimes, these self-deformations, or gaits, produce center-of-mass displacements that are independent of the period of the gait, due to the dominance of dissipation over inertia \cite{Purcell1977}. Once thought to be restricted to the microscopic domain of low Reynolds number swimming in water and complex biofluids, subsequent work has revealed that many terrestrial locomotor systems, such as snakes \cite{hu2009mechanics} and centipedes \cite{chong2023self}, also operate in overdamped regimes where inertia, and hence coasting (defined as the ratio of the time taken to stop after cessation of activity to the cycle duration) is negligible  \cite{Rieser2019, lauga2009hydrodynamics}. Principles of locomotion in overdamped regimes can therefore help to describe organisms across scales, environments, and taxa \footnote{Similar principles, often in the form of relationships between gait parameters have been previously identified in many different categories of locomotion, including bipedal, quadrupedal, hexapodal and myriapodal locomotion \cite{Heglund1974,Hildebrand1989-rb,Full1991,Chong2023}, flight \cite{Sullivan2019}, and aquatic swimming \cite{Gazzola2014, Sanchez-Rodriguez2023}}.

Because the displacement is rate-independent, theoretical models of low-coasting locomotion (e.g. resistive force theory \cite{gray1955propulsion,Zhang2014-fi}, slender body theory \cite{batchelor1970slender}, geometric mechanics \cite{Shapere1989,Rieser2024}) often describe relationships between spatial variables. On the other hand, temporal parameters, such as the frequency of undulation $\omega$, are nonetheless important; they affect the energetics of locomotion, and organisms operate under power constraints. Understanding why organisms in overdamped regimes select particular gaits requires consideration of the tradeoffs between optimal kinematics and power constraints \cite{Schiebel2020mitigating}. These tradeoffs indirectly couple the spatial and temporal parameters of the gait.

Here we identify a link between the temporal and spatial traveling wave dynamics of undulatory gaits in the form of a functional relationship between the wavenumber $k$ and frequency $\omega$. We first observed this relationship in an experimental study of nematode locomotion in a variety of complex environments. We explain our observations by deriving an active `dispersion relation' $\omega(k)$ from force balance, \textcolor{black}{assuming a muscle torque in the form of a traveling wave phase shifted with respect to the body bending wave \cite{mcmillen2008nonlinear, ding2013emergence, pierce2025neuromechanical}.} This dispersion relation holds not only for worms moving through a broad range of rheologies, but also for a set of unrelated undulatory systems (spermatozoa, and fish larvae).

\begin{figure}[ht]
  \includegraphics[width=\columnwidth]{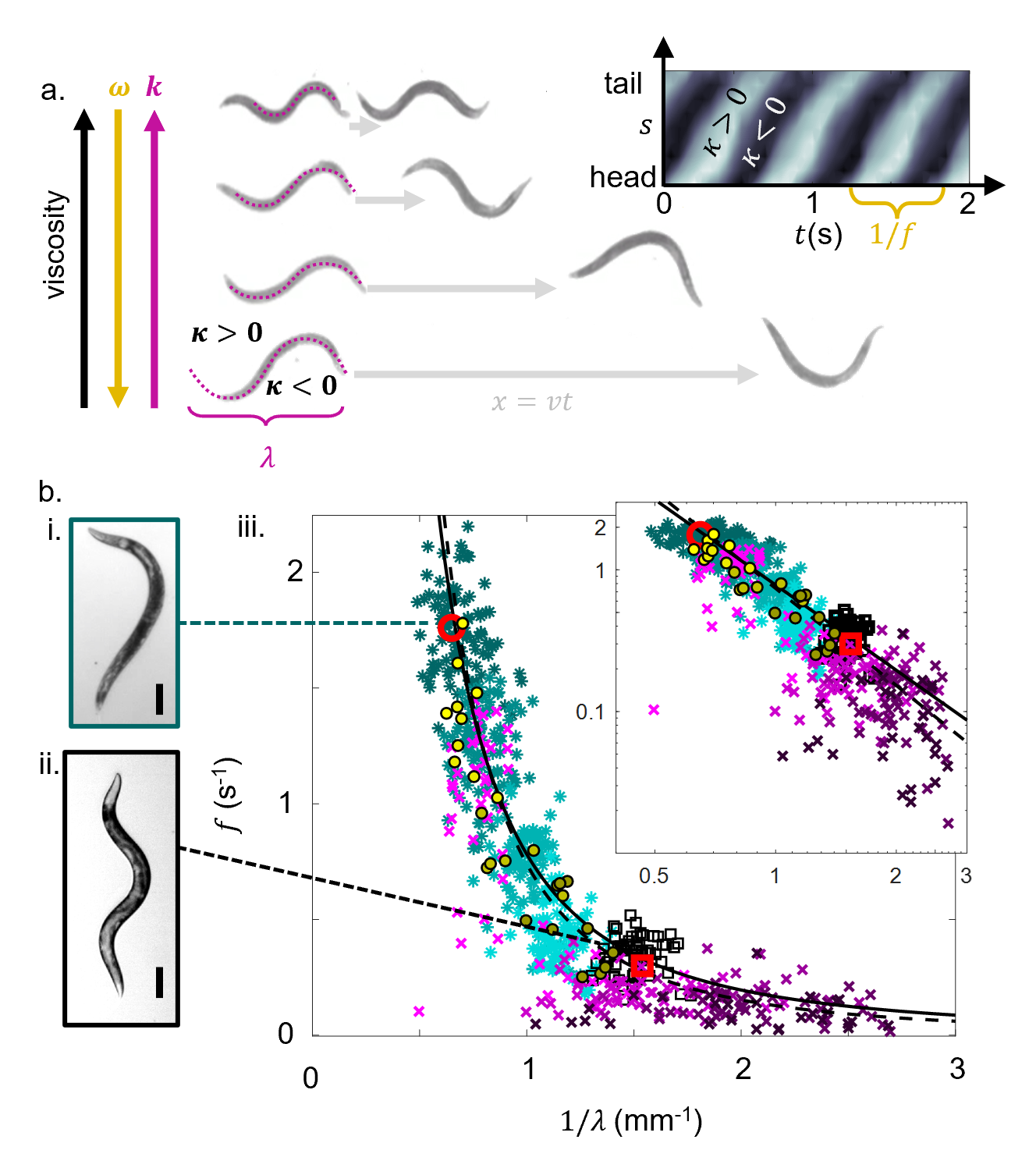}
       \caption{\label{fig:Fig_worm}  
       Dispersion relation showing the approximate inverse quadratic scaling of $\omega \propto f$ with $k\propto 1/\lambda $ for %the nematode worm 
       \textit{C. elegans} under changing environmental rheology. (a) Brightfield images of worm body shapes traveling through fluids of different viscosities, final positions are after 4.5s elapsed time. Violet curves illustrate full wave periods to show the changing wavelength $\lambda$. Inset, heatmap showing curvature $\kappa$ (in units of $mm^{-1}$) along the body coordinate $s$ (the arc length from head to tail in units of body length) and through time showing alternating bands of positive and negative curvature oscillating at a frequency $f$. (b) Postures of nematodes in buffer (b, i) and agar (b, ii) (scale bar, 1 mm). Linear (b, iii) and log plot (b, iii, inset) of the experimental dispersion relation for nematodes in diverse environments shown with the inverse quadratic (solid black line) and power law (dashed black line) fits, along with experimental data \textcolor{cyan}{$\mathbf{\boldsymbol{*}}$} methylcellulose (0-3\%), $\square$ Agar, \textcolor{magenta}{$\boldsymbol{\times}$} PEG (1-5\%), \textcolor{yellow}{$\bullet$} dextran from Butler et al \cite{butler2015consistent}, \textcolor{red}{$\bigcirc$} buffer and \textcolor{red}{$\boldsymbol{\square}$} agar from Fang-yen et al \cite{fang2010biomechanical}. Power law exponent is $-2.28\pm0.30$.}
       
\end{figure}

\paragraph{Experimental Observations} We begin by describing a systematic experimental study of the forward swimming gaits of the nematode \textit{Caenorhabditis elegans}. \textit{C. elegans} encounters diverse environments in its native habitat \cite{felix2010natural} and can locomote in a variety of complex laboratory environments via dorso-ventral bends that pass from head to tail \cite{juarez2010motility,shen2011undulatory,majmudar2012experiments, Wang2023mechanical}. This species is known to modify its gait parameters as a function of environmental viscosity to maintain gait performance \cite{fang2010biomechanical}. See Fig. \ref{fig:Fig_worm}. We experimentally measured $\omega$ and $k$ \footnote{To avoid confusion, we use the following notational convention throughout the manuscript. For dimensionless quantities, we use the angular frequency and wavenumber $\omega$ and $k$. When presenting experimental data, we use the corresponding linear frequency and inverse wavelength in SI units, denoted $f$ and $1/\lambda$, respectively.} for \textit{C. elegans} in a diverse set of environments including buffer solutions, methylcellulose mixtures of different viscosities, %(weakly viscoelastic fluids)
polyethylene glycol (PEG) hydrogels with a range of bulk moduli, and agar surfaces. (See the SI for details.) We also compared our measured gaits to two literature sources, including previously measured buffer swimming and agar crawling gaits \cite{fang2010biomechanical}, and swimming gaits in Dextran mixtures with various viscosities \cite{butler2015consistent}. % which remain Newtonian over a broad range of concentrations. 
Surprisingly, across these diverse rheologies, \textit{C. elegans'} gaits fall approximately on a single curve given by the dispersion relation $\omega(k) \propto k^{-2}$. Figure \ref{fig:Fig_worm}(b) shows the data along with a fit to an inverse quadratic function (solid line, $R^2=0.98$) and a power law fit (dashed line, $R^2=0.99$), which yielded an exponent of $-2.28\pm0.30$.  

\begin{figure}[ht]
    \includegraphics[width=0.475\textwidth]{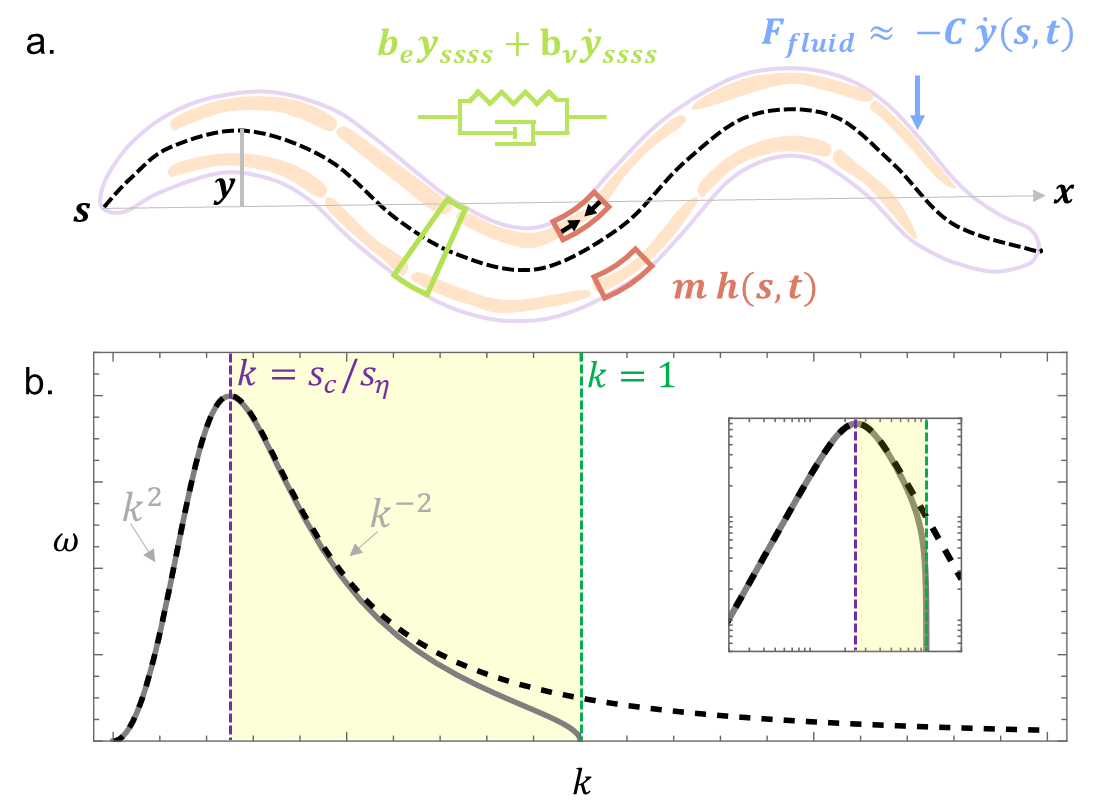}
    \caption{\label{fig:fig_model}
   An active damped beam model and resulting dispersion relations. (a) Model variables and internal and external forces for a generic undulator in a fluid. (b) The real part of the full dispersion relation (dashed, black curve) and the real part of the dispersion curve with the constraint of wave stability imposed (solid, gray curve).  (b, inset) The same curves on a log-log scale. Dashed vertical lines indicate the value of $k$ separating quadratic and inverse scaling (purple) and the wave stability cutoff (green), for $s_c/s_\eta = 0.25$. Yellow shaded region shows the domain in $k$ which applies to the locomotion of nematodes.}
\end{figure}

\paragraph{Deriving the Model} To explain this observation, we construct a mechanical model based on force balance. We begin by considering a driven viscoelastic (Kelvin-Voight) beam previously used to model terrestrial undulation of snakes on frictional surfaces (where friction dominates inertia) \cite{Guo2008}. This model was subsequently used to describe nematodes immersed in a low-Reynolds number fluid \cite{fang2010biomechanical}. The linear force balance along the body is given by
    
\begin{equation}\label{wormeom}
    b_e y_{ssss} + b_{\eta} \dot{y}_{ssss} + mh_{ss} = -C \dot{y},
\end{equation}

\noindent which equates internal and external linear force densities. Here $y(s,t)$ is the lateral displacement of the beam, $s$ is the arclength, and $t$ is time. See Fig. \ref{fig:fig_model}(a). The force densities are $b_e y_{ssss}$ for the elastic body force, $b_{\eta} \dot{y}_{ssss}$ for the viscous body force,  and $-C \dot{y}$ for the fluid drag force in the $y$-direction \footnote{In the small amplitude limit used in this manuscript, the normal fluid drag is approximated by  $C \dot{y}$ and the $y$-component of the tangential fluid drag is negligible.}. Here $mh_{ss}$ is the force density from the active moment produced by the muscles, which we have written in terms of a constant $m$ with units of torque, and a dimensionless function $h(s,t)$, which describes the spatiotemporal variation of the driving torque. In this equation, the variable $s$ in the subscript denotes partial derivative with respect to $x$ (using the approximation $s\approx x$) and the overdot denotes partial derivative with respect to time. Henceforth $y(s,t) = y(x,t)$. 
 
This force balance equation (\ref{wormeom}) implies two characteristic length scales and a single characteristic time scale. (See the SI for a derivation.) Here we have chosen to scale the lengths $y$ and $s$ by the length scale $s_c = b_e/m$, which represents the minimal radius of curvature achieved by the body when the muscle torque $m$ is balanced entirely by the body elasticity, and the time by $t_c = b_{\eta}/b_e$, which represents the viscoelastic relaxation time of the passive body. The second length scale implied by (\ref{wormeom}), $s_{\eta}=(\frac{b_{\eta}}{C})^{1/4}$, relates to the relative strength of internal and external dissipation terms $b_{\eta} \dot{y}_{ssss}$ and $-C\dot{y}$. Specifically,  $\lambda = s_{\eta}$ is the wavelength at which the dissipative terms are equal in magnitude. While we have not chosen to scale the equation of motion using $s_{\eta}$, it remains important in characterizing the dynamics, as shown below.

\textcolor{black}{Using $s_c$ and $t_c$, we produce the following dimensionless equation of motion.} %(See the SI for a full derivation.)}

\textcolor{black}{
\begin{equation}    \label{adimwormeom}\upsilon_{\sigma\sigma\sigma\sigma}+\dot{\upsilon}_{\sigma\sigma\sigma\sigma}+h_{\sigma\sigma}=-\left(\frac{s_c}{s_\eta}\right)^4\dot{\upsilon}.
\end{equation}}

\noindent
\textcolor{black}{Here $\upsilon$ represents the nondimensional variable that corresponds to $y$. The dimensionless independent variables $\sigma$ and $\tau$ correspond to $s$ and $t$, respectively. The overdot now represents partial differentiation with respect to $\tau$. The function $h(\sigma,\tau)$ represents the spatiotemporal variation in the muscle torque, as above, but as a function of the new dimensionless variables. }

\textcolor{black}{
We solve (\ref{adimwormeom}) by substituting a right-traveling plane wave ansatz solution
\[ \upsilon(\sigma, \tau) = \exp(i(k\sigma - \omega \tau)),\]
with wavenumber $k$ and angular frequency $\omega$ and by specifying a particular form of the muscle torque function $h$. %As noted above, 
Previous measurements in fish, lizards, and nematodes showed that muscle activity patterns are approximately sinusoidal and are temporally phase-shifted relative to the body bends (referred to as neuromechanical phase lags \cite{mcmillen2008nonlinear, ding2013emergence, pierce2025neuromechanical}). For this reason, we may write}
\textcolor{black}{
\begin{equation}\label{forceterm}
    h=e^{-i\phi}\cdot \upsilon.
\end{equation}}

\noindent\textcolor{black}{Here $\phi>0$ represents the phase lag relative to the body bending waves. Finally, we solve for $\omega$ to produce the  dispersion relation with real and imaginary parts}

\textcolor{black}{
\begin{equation}\label{re}
    Re\{\omega (k)\} =\frac{k^2\sin(\phi)}{(s_c/s_{\eta})^4 +k^4}, 
\end{equation}}

\noindent\textcolor{black}{and}

\textcolor{black}{\begin{equation}\label{im}
     Im\{\omega (k)\} = \frac{k^2 \cos(\phi)-k^4}{(s_c/s_{\eta})^4 +k^4}.
\end{equation}}

\textcolor{black}{The oscillation of a solution to (\ref{wormeom}) is governed by the real part of the dispersion relation (\ref{re}), which is plotted in Fig. \ref{fig:fig_model} (b) as a dashed curve. For oscillations to persist in time, the imaginary part (\ref{im}) must vanish. Setting (\ref{im}) to zero, we establish the following criterion for wave stability}
\textcolor{black}{
\begin{equation}\label{stabilityCriterion}
    k^2=\cos\phi.
\end{equation}}

\noindent \textcolor{black}{We enforce this criterion by substituting (\ref{stabilityCriterion}) into (\ref{re}), yielding}

\textcolor{black}{
\begin{equation}\label{re_stability}
        Re\{\omega (k)\} =\frac{k^2 \sqrt{1-k^4}}{(s_c/s_{\eta})^4 +k^4}, 
\end{equation}}

\noindent\textcolor{black}{which introduces an additional factor of $\sqrt{1-k^4}$ to (\ref{re}). 
This modified dispersion curve %, after incorporating the wave stability criterion, 
is shown in Fig. \ref{fig:fig_model}(b) as a solid, gray curve. The stability criterion places a constraint, $k<1$, on the possible values of the dimensionless wavenumber. Because of our choice of nondimensionalization, this is equivalent to the condition that the wavelength be greater than the characteristic lengthscale, $\lambda>s_c$ (in natural units).}

\textcolor{black}{\paragraph{Two Scaling Regimes} 
We now consider two scaling regimes in the dispersion relation (\ref{re_stability}).  We first consider the case where $k > s_c/s_\eta$ and $k < 1$. In this domain, we recover the experimentally observed relation $\omega\propto k^{-2}$, as illustrated in Fig. 2 (b). This implies that $k^{-2}$ can only be observed when the wavelength $\lambda/2\pi$, taken in natural units, is less than $s_{\eta}$, but greater than $s_c$.  This, in turn, suggests that nematodes operate in a regime where the largest source of dissipation is within the body's bending degree of freedom, and not from the surrounding fluid, since $\lambda<s_{\eta}$ is the regime where internal dissipation exceeds external viscous dissipation forces. This finding suggests that internal damping within the body plays a more significant role in nematode biomechanics than previously thought \cite{fang2010biomechanical}.}

\textcolor{black}{To model this internal dissipation-dominated regime, we eliminate the viscous drag term in (\ref{adimwormeom}), producing a simplified equation of motion. %(See SI.) 
Calculating the dispersion relation for this regime, we recover the same stability criterion described above (\ref{stabilityCriterion}). Substituting into the real part of the dispersion relation, we obtain}

\textcolor{black}{\begin{equation}\label{invquad_coeff}
    Re\{\omega(k)\}=\frac{\sqrt{1-k^4}}{k^2}.
\end{equation}}

\noindent\textcolor{black}{By restoring units, we can determine the constant of proportionality (to leading order)}

\textcolor{black}{\begin{equation}
    f=\frac{\lambda^2}{2\pi t_c s_c^2} = \alpha\lambda^2.
\end{equation}}

\noindent \textcolor{black}{Here we have defined the constant $\alpha = (2\pi t_cs_c^2)^{-1}=m^2/(2\pi b_{\eta}b_e)$. From a fit to the data in Fig. 1 (a), we estimate the value of the proportionality constant to be $\alpha= 0.69 \pm 0.02 s^{-1}$mm$^{-2}$.}%We emphasize that this proportion holds for a wide range of rheologies.

\textcolor{black}{The second case we consider is the regime where $k < s_c/s_{\eta}$ and $k<1$. Here, equation (\ref{re_stability}) produces $\omega \propto {k}^2$ and the dominant dissipation term is the external viscous dissipation. In a similar way we may recover this scaling by eliminating the internal dissipation term from (\ref{adimwormeom}), which yields }

\textcolor{black}{\begin{equation}\label{quad_coeff}
    Re\{\omega(k)\} = \frac{s_\eta^2}{s_c^2}k^2\sqrt{1-k^4},
\end{equation}}

\noindent \textcolor{black}{after enforcing the stability criterion \footnote{The full equation of motion and both limiting cases produce the same wave stability criterion.}. Restoring units, we find, to leading order}

\textcolor{black}{
\begin{equation}\label{propforfish}
    f=\frac{2\pi s_\eta^2}{t_c\lambda^2}=\frac{\beta}{\lambda^2},
\end{equation}}

\noindent \textcolor{black}{where we have defined $\beta=2\pi s_\eta^2/t_c=2\pi b_e (b_\eta C)^{-1/2}$.}

\textcolor{black}{ \paragraph{Boundary Conditions} The analysis above does not consider the finite size of the organisms. In effect, we have ignored boundary conditions on (\ref{adimwormeom}) by assuming the solutions are defined on an infinite domain. Previous work in low-coasting regimes has neglected the effect of the boundary \cite{fang2010biomechanical}. In the case of snake locomotion, Guo et al \cite{Guo2008} ignored end effects, instead modeling undulations with periodic boundary conditions. They justified this approach by noting that the length of the organism was much larger than one undulation wavelength. This criterion may apply to some organisms other than snakes, but certainly not to organisms with smaller wavelength to body length ratios, including nematodes.}

\textcolor{black}{We re-derived the dispersion relation after applying force and moment free boundary conditions \cite{McMillen2006elastic}, using the so-called ``recoil correction" method \cite{Lighthill1960note,Hess1984fast,Cheng1998analysis}, which is frequently used in the context of inertial aquatic swimming. (See SI for derivation and discussion.) Interestingly, we recovered the inverse quadratic dispersion relation for the equation of motion (\ref{wormeom}), along with the stability criterion (\ref{stabilityCriterion}). The recoil corrected solutions did not support a quadratic dispersion relationship, however. We will return to this point later in the discussion of experimental evidence for a quadratic scaling regime.}

\begin{figure}[h]
    \centering
    \includegraphics[width=3in]{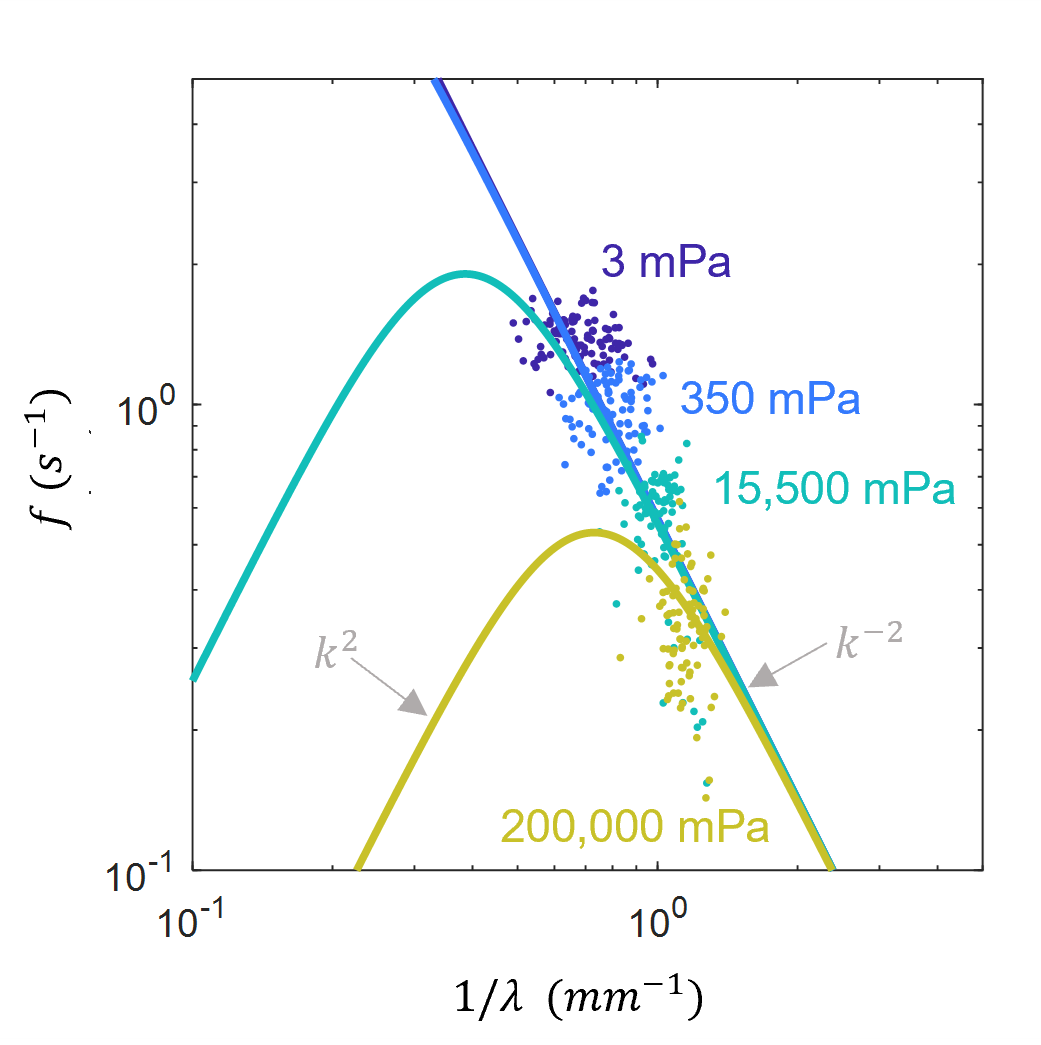}
    \caption{\label{fig:Fig_visc}     
    Dispersion relation (\ref{re}) in natural units for different values of fluid drag coefficient $C$ taken from methylcellulose data. All the \textit{C. elegans} data lie in the $k>s_c/s_{\eta}$ regime and are insensitive to the precise location of the peak at $k = s_c/s_{\eta}$. }
\end{figure}

\paragraph{Dependence on Model Parameters} Having identified the scaling regimes in the dispersion relation, we proceed to discuss how the model parameters change with the environment, and we return to our experimental data. For the nematodes to obey a constant scaling relationship across individuals and across the various experimental environments, $\alpha$ must be constant. The lack of dependence on $C$ in equation (\ref{invquad_coeff}) suggests that viscosity shifts will not alter the scaling, but only change the location of the maximum value separating the $k^2$ and $k^{-2}$ regimes, because $s_\eta$ (and hence the value of $k$ that separates the two regimes) does depend on viscosity. Restoring units to the model using measured viscosities and  comparing with the  methylcellulose data, Fig. \ref{fig:Fig_visc} shows that, indeed, while the viscosity affects the relative distance of the experimental wavenumbers to the peak of the dispersion curve, it does not change the overall $k^{-2}$ relationship. Increasing the viscosity does, however, induce the worm to select a different $\omega$ as previously noted \cite{fang2010biomechanical}.

We next consider changes in the \textit{elasticity} of the surrounding medium. For example, in the PEG hydrogel experiments [magenta points in Fig. \ref{fig:Fig_worm}(b)], nematodes encounter a highly elastic environment relative to methylcellulose, dextran and buffer solutions. In these data, the scaling persists because the addition of a linear elastic term to (\ref{adimwormeom}) does not impact the scaling of the real part of the dispersion (\ref{re}), and only adds a constant offset to the numerator in (\ref{im}). Hence, environmental elasticity will change the stability criterion, but does not impact the scaling of the real part of the dispersion for values of $k$ that maintain stability. 

\paragraph{Experimental Evidence for $k^2$ Scaling} Finally, we asked if any organisms might follow the $k^2$ dispersion relation predicted by the model when $k< s_c/s_{\eta}$ and $k<1$. \textcolor{black}{While the application of boundary conditions theoretically precludes the $k^2$ scaling within our model, as previously noted, organisms, such as snakes, with sufficiently small wavelength to body length ratios can be modeled without taking into account end effects \cite{Guo2008}}. We therefore considered several literature sources of gait parameters measured in other undulators, a meta-analysis of fish swimming \cite{Van_Weerden2014}, the bank of swimming organisms at the micron scale (BOSO-Micro) database \cite{Velho_Rodrigues2021}, where we investigated the gaits of spermatozoa, an analysis of other nematode species' gaits \cite{Gray1964locomotion}, and a study of polychaete worms in water and %in 
sediment \cite{Dorgan2013}. The non-\textit{C. elegans} nematodes and polychaete worms displayed frequencies that \textit{decreased} with $k$ across environments. (The data was insufficient to evaluate the scaling, however.) In contrast, for both the spermatozoa and the fish data, $\omega$ \textit{increased} with $k$, suggesting that they may operate in the external dissipation-dominated regime implied by our model. 

\begin{figure}[htbp]
\includegraphics[width=\columnwidth]{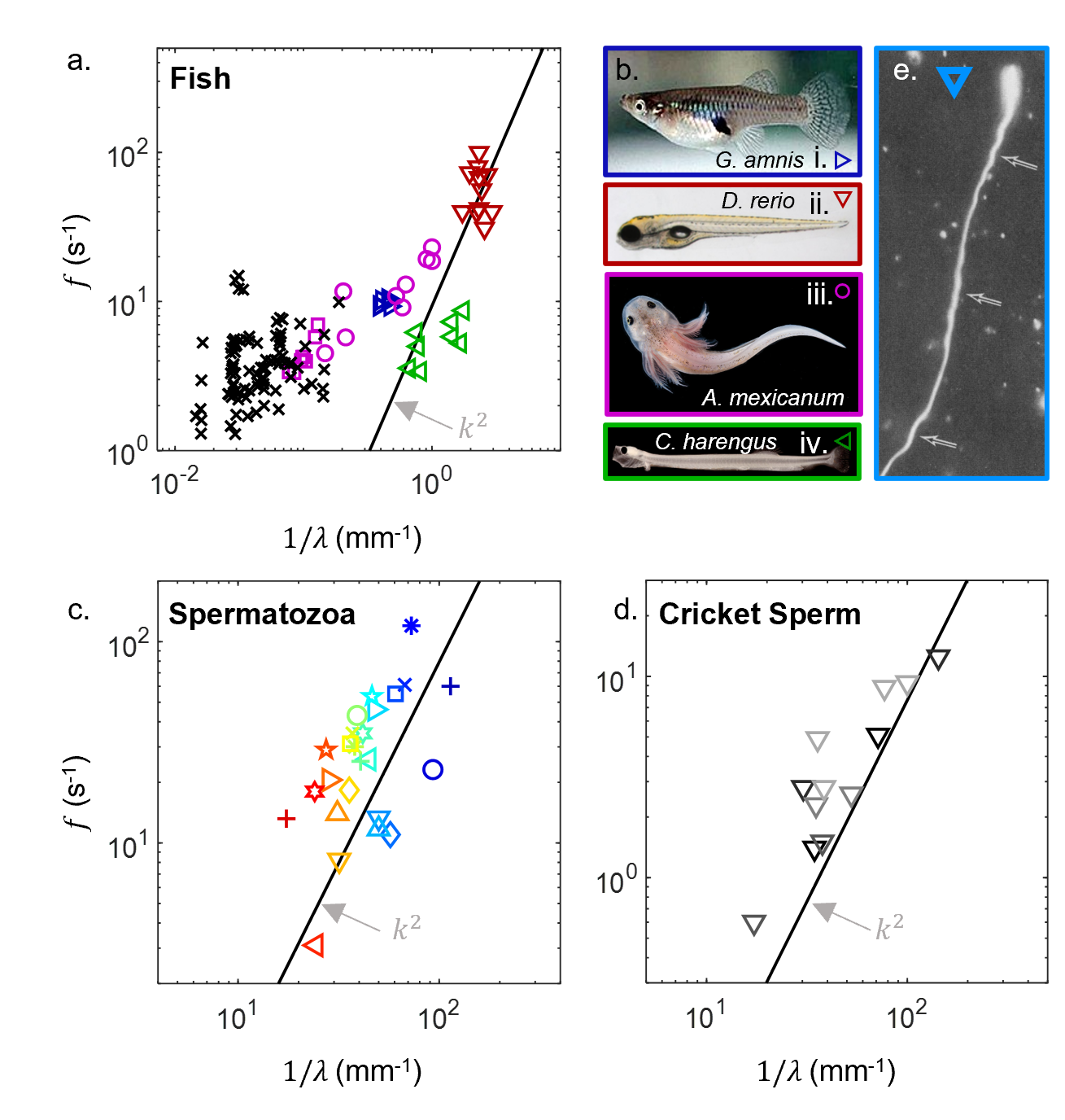}
\caption{\label{fig:Fig_larvae_sperm} 
%In contrast to nematode worms, 
Fish larvae and spermatozoa display dispersion relations where $f$ is an increasing function of $1/\lambda$, indicating that external fluid dissipation dominates internal dissipation. (a)  Data from fish swimming meta-analysis with selected low-inertia swimmers separated by species (high-inertia swimmers plotted in black).  Images of a mosquitofish \textit{G. amnis} ($\sim$cm, Wikipedia) (b, i), zebrafish larva, \textit{D. rerio} ($\sim$mm from \cite{Maes2012}) (b, ii), an axolotl larva \textit{A. mexicanum} ($\sim$cm photo credit John P. Clare) (b,iii), and a herring larva \textit{C. harengus} ($\sim$mm from \cite{Fischbach2023}) (b, vi). (c) Spermatozoa data from different species taken from \cite{Velho_Rodrigues2021} along with quadratic fit, 
\textcolor{sperm1}{$\mathbf{\boldsymbol{+}}$}  \textit{P. maxima} (turbot),
\textcolor{sperm2}{$\mathbf{\boldsymbol{\circ}}$} \textit{A. curtula} (beetle),
\textcolor{sperm3}{$\mathbf{\boldsymbol{*}}$} \textit{Lygaeus} (milkweed bug),
\textcolor{sperm4}{$\mathbf{\boldsymbol{\times}}$} \textit{T. thynnus} (tuna),
\textcolor{sperm5}{$\mathbf{\boldsymbol{\square}}$} \textit{M. merluccius} (hake),
\textcolor{sperm6}{$\mathbf{\boldsymbol{\diamond}}$} \textit{C. capitata} (fly),
\textcolor{sperm7} {$\mathbf{\boldsymbol{\triangledown}}$} Cricket,
\textcolor{sperm8}{$\mathbf{\boldsymbol{\triangle}}$} \textit{B. marinus} (toad),  
\textcolor{sperm9}{$\mathbf{\boldsymbol{\triangleright}}$} \textit{Colobocentrotus} (sea urchin),
\textcolor{sperm10}{$\mathbf{\boldsymbol{\star}}$} \textit{G. morhua} (cod),
\textcolor{sperm11}{$\mathbf{\boldsymbol{\triangleleft}}$} \textit{Chaetopterus} (annelid),
\textcolor{sperm12}{$\mathbf{\boldsymbol{\davidsstar}}$} \textit{Psammechinus} (sea urchin),
\textcolor{sperm13}{$\mathbf{\boldsymbol{+}}$} \textit{G. domesticus} (domestic fowl),
\textcolor{sperm14}{$\mathbf{\boldsymbol{\circ}}$} \textit{Ostrea} (oyster),
\textcolor{sperm15}{$\mathbf{\boldsymbol{*}}$} \textit{Lytechinus} (sea urchin),
\textcolor{sperm16}{$\mathbf{\boldsymbol{\times}}$} \textit{Ciona} (tunicate),
\textcolor{sperm17}{$\mathbf{\boldsymbol{\square}}$} \textit{S. purpuratus} (sea urchin),
\textcolor{sperm18}{$\mathbf{\boldsymbol{\diamond}}$} \textit{Myzostomus} (worm),
\textcolor{sperm19}{$\mathbf{\boldsymbol{\triangledown}}$} \textit{C. mellus} (midge),
\textcolor{sperm20}{$\mathbf{\boldsymbol{\triangle}}$} \textit{Homo} (human),
\textcolor{sperm21}{$\mathbf{\boldsymbol{\triangleright}}$} \textit{Bos}, bull,
\textcolor{sperm22}{$\mathbf{\boldsymbol{\star}}$} \textit{Ovis} (ram), 
\textcolor{sperm23}{$\mathbf{\boldsymbol{\triangleleft}}$} \textit{M. sclaris} (fly),
\textcolor{sperm24}{$\mathbf{\boldsymbol{\davidsstar}}$} Rabbit,
\textcolor{sperm24}{$\mathbf{\boldsymbol{+}}$} \textit{Mus} (mouse),
(d) Cricket sperm dispersion relation for multiple individuals with multiple measurements along each flagellum from \cite{Rikmenspoel1978} (different gray values represent different individuals), along with (e) an image of multiple waves on a single cricket flagellum ($\sim0.1-1$mm)\cite{Rikmenspoel1978}.}
\end{figure}

For many fish, the body and fluid dynamics are inertial. Unsurprisingly, the majority of the data in the meta-analysis \cite{Van_Weerden2014} do not fit well to a $k^2$ model, which assumes that inertia is negligible. A small number of organisms, however, with Reynolds numbers (Re) under 2,000 and Strouhal numbers (St) greater than 0.7 approximately fit to a $k^2$ dispersion relation [see Fig. \ref{fig:Fig_larvae_sperm} (a)]. We note that while Re $\sim2,000$ is inertial, the drag coefficient remains approximately linear in velocity at intermediate Reynolds numbers (See for example \cite{PantonHydro}, Ch. 14). Furthermore, Ref. \cite{Van_Weerden2014} noted the existence of two distinct swimming regimes based on the same criteria. % we have employed here.
Unsurprisingly, organisms in this regime were some of the smallest species in the study, and with one exception were collected from different larval stages [see Fig. \ref{fig:Fig_larvae_sperm}(a)]. We estimate the proportionality constant in  (\ref{propforfish}) to be $\beta_{Fish} = 9.4 \pm 1.0 s^{-1}m^{2}$. \textcolor{black}{For many fish larvae, end effects cannot be safely ignored. Hence, the effect of the boundary requires further investigation to account for the observed scaling.}

We also fit the spread of the spermatozoa data to a $k^2$ relationship and find a rough agreement, with a constant of $\beta_{Sp.} = (8 \pm 3)\times 10^{-3} s^{-1}mm^{2}$. Cricket sperm are particularly interesting in terms of their dispersion relations, because their unusually long flagella ($\approx mm$) exhibit spatial variation in $k$ and $\omega$  \cite{Rikmenspoel1978}. Thus, an individual cell's flagellum allows a test of the model where the biomechanical parameters are likely held constant \textcolor{black}{and end effects may be negligible.} Figure \ref{fig:Fig_larvae_sperm}(d) shows cricket sperm data from \cite{Rikmenspoel1978} along with a fit yielding a constant of $\beta_{Crick.} = (7 \pm 4)\times 10^{-5} s^{-1}mm^{2}$.

\paragraph{Summary and Conclusion} 

We discovered a previously unknown relationship between the wavenumber and undulation frequencies of nematode gaits that persists across diverse environmental rheologies. We explained the origin of this relationship with a simple mechanical model. Surprisingly, our model also captures the approximate scaling of the gait parameters of larval aquatic swimmers and spermatozoa.

Because $\omega$ and $k$ are functionally related, selecting one variable determines the other. Our model does not explain how the undulator's locomotor control system selects a point along the dispersion curve. For nematodes, based on models of proprioceptive feedback \cite{Ji2021}, we hypothesize that the motor circuit controls $\omega$, decreasing it in response to increasing external viscosity \cite{fang2010biomechanical}. Consequently, $k$ is either fixed spontaneously through mechanics or as a result of mechanical entrainment of coupled motor neuron oscillations \cite{johnson2021neuromechanical}. Nematodes target a set of gaits that maintain a relatively constant overall speed across different environments \cite{fang2010biomechanical}. Higher wavenumber gaits produce higher kinematic efficiencies \cite{pierce2025neuromechanical}, which compensate for the reduction in undulation frequency. Our results imply that this compensatory gait adaptation is \textit{mechanically} enforced through the dispersion relation.

\section{Acknowledgements}
This work was supported by  NIH R01AG082039,  NSF Physics of Living Systems Student Research Network (GR10003305), NSF-Simons Southeast Center for Mathematics and Biology (SCMB) through the National Science Foundation grant DMS1764406 and Simons Foundation/SFARI  grant 594594. We would like to thank Prof. Andrew Zangwill for useful discussions. We would further like to thank the editor and reviewers for their insightful comments during the review process.
\providecommand{\noopsort}[1]{}\providecommand{\singleletter}[1]{#1}%\providecommand{\noopsort}[1]{}\providecommand{\singleletter}[1]{#1}%


\begin{thebibliography}{47}%
\makeatletter
\providecommand \@ifxundefined [1]{%
 \@ifx{#1\undefined}
}%
\providecommand \@ifnum [1]{%
 \ifnum #1\expandafter \@firstoftwo
 \else \expandafter \@secondoftwo
 \fi
}%
\providecommand \@ifx [1]{%
 \ifx #1\expandafter \@firstoftwo
 \else \expandafter \@secondoftwo
 \fi
}%
\providecommand \natexlab [1]{#1}%
\providecommand \enquote  [1]{``#1''}%
\providecommand \bibnamefont  [1]{#1}%
\providecommand \bibfnamefont [1]{#1}%
\providecommand \citenamefont [1]{#1}%
\providecommand \href@noop [0]{\@secondoftwo}%
\providecommand \href [0]{\begingroup \@sanitize@url \@href}%
\providecommand \@href[1]{\@@startlink{#1}\@@href}%
\providecommand \@@href[1]{\endgroup#1\@@endlink}%
\providecommand \@sanitize@url [0]{\catcode `\\12\catcode `\$12\catcode
  `\&12\catcode `\#12\catcode `\^12\catcode `\_12\catcode `\%12\relax}%
\providecommand \@@startlink[1]{}%
\providecommand \@@endlink[0]{}%
\providecommand \url  [0]{\begingroup\@sanitize@url \@url }%
\providecommand \@url [1]{\endgroup\@href {#1}{\urlprefix }}%
\providecommand \urlprefix  [0]{URL }%
\providecommand \Eprint [0]{\href }%
\providecommand \doibase [0]{http://dx.doi.org/}%
\providecommand \selectlanguage [0]{\@gobble}%
\providecommand \bibinfo  [0]{\@secondoftwo}%
\providecommand \bibfield  [0]{\@secondoftwo}%
\providecommand \translation [1]{[#1]}%
\providecommand \BibitemOpen [0]{}%
\providecommand \bibitemStop [0]{}%
\providecommand \bibitemNoStop [0]{.\EOS\space}%
\providecommand \EOS [0]{\spacefactor3000\relax}%
\providecommand \BibitemShut  [1]{\csname bibitem#1\endcsname}%
\let\auto@bib@innerbib\@empty
%</preamble>
\bibitem [{\citenamefont {Purcell}(1977)}]{Purcell1977}%
  \BibitemOpen
  \bibfield  {author} {\bibinfo {author} {\bibfnamefont {E.~M.}\ \bibnamefont
  {Purcell}},\ }\href
  {https://pubs.aip.org/aapt/ajp/article-pdf/45/1/3/11809839/3_1_online.pdf}
  {\bibfield  {journal} {\bibinfo  {journal} {Am. J. Phys.}\ }\textbf {\bibinfo
  {volume} {45}},\ \bibinfo {pages} {3} (\bibinfo {year} {1977})}\BibitemShut
  {NoStop}%
\bibitem [{\citenamefont {Hu}\ \emph {et~al.}(2009)\citenamefont {Hu},
  \citenamefont {Nirody}, \citenamefont {Scott},\ and\ \citenamefont
  {Shelley}}]{hu2009mechanics}%
  \BibitemOpen
  \bibfield  {author} {\bibinfo {author} {\bibfnamefont {D.~L.}\ \bibnamefont
  {Hu}}, \bibinfo {author} {\bibfnamefont {J.}~\bibnamefont {Nirody}}, \bibinfo
  {author} {\bibfnamefont {T.}~\bibnamefont {Scott}}, \ and\ \bibinfo {author}
  {\bibfnamefont {M.~J.}\ \bibnamefont {Shelley}},\ }\href@noop {} {\bibfield
  {journal} {\bibinfo  {journal} {Proceedings of the National Academy of
  Sciences}\ }\textbf {\bibinfo {volume} {106}},\ \bibinfo {pages} {10081}
  (\bibinfo {year} {2009})}\BibitemShut {NoStop}%
\bibitem [{\citenamefont {Chong}\ \emph
  {et~al.}(2023{\natexlab{a}})\citenamefont {Chong}, \citenamefont {He},
  \citenamefont {Li}, \citenamefont {Erickson}, \citenamefont {Diaz},
  \citenamefont {Wang}, \citenamefont {Soto},\ and\ \citenamefont
  {Goldman}}]{chong2023self}%
  \BibitemOpen
  \bibfield  {author} {\bibinfo {author} {\bibfnamefont {B.}~\bibnamefont
  {Chong}}, \bibinfo {author} {\bibfnamefont {J.}~\bibnamefont {He}}, \bibinfo
  {author} {\bibfnamefont {S.}~\bibnamefont {Li}}, \bibinfo {author}
  {\bibfnamefont {E.}~\bibnamefont {Erickson}}, \bibinfo {author}
  {\bibfnamefont {K.}~\bibnamefont {Diaz}}, \bibinfo {author} {\bibfnamefont
  {T.}~\bibnamefont {Wang}}, \bibinfo {author} {\bibfnamefont {D.}~\bibnamefont
  {Soto}}, \ and\ \bibinfo {author} {\bibfnamefont {D.~I.}\ \bibnamefont
  {Goldman}},\ }\href@noop {} {\bibfield  {journal} {\bibinfo  {journal}
  {Proceedings of the National Academy of Sciences}\ }\textbf {\bibinfo
  {volume} {120}},\ \bibinfo {pages} {e2213698120} (\bibinfo {year}
  {2023}{\natexlab{a}})}\BibitemShut {NoStop}%
\bibitem [{\citenamefont {Rieser}\ \emph {et~al.}(2019)\citenamefont {Rieser},
  \citenamefont {Chong}, \citenamefont {Gong}, \citenamefont {Astley},
  \citenamefont {Schiebel}, \citenamefont {Diaz}, \citenamefont {Pierce},
  \citenamefont {Lu}, \citenamefont {Hatton}, \citenamefont {Choset},\ and\
  \citenamefont {Goldman}}]{Rieser2019}%
  \BibitemOpen
  \bibfield  {author} {\bibinfo {author} {\bibfnamefont {J.~M.}\ \bibnamefont
  {Rieser}}, \bibinfo {author} {\bibfnamefont {B.}~\bibnamefont {Chong}},
  \bibinfo {author} {\bibfnamefont {C.}~\bibnamefont {Gong}}, \bibinfo {author}
  {\bibfnamefont {H.~C.}\ \bibnamefont {Astley}}, \bibinfo {author}
  {\bibfnamefont {P.~E.}\ \bibnamefont {Schiebel}}, \bibinfo {author}
  {\bibfnamefont {K.}~\bibnamefont {Diaz}}, \bibinfo {author} {\bibfnamefont
  {C.}~\bibnamefont {Pierce}}, \bibinfo {author} {\bibfnamefont
  {H.}~\bibnamefont {Lu}}, \bibinfo {author} {\bibfnamefont {R.~L.}\
  \bibnamefont {Hatton}}, \bibinfo {author} {\bibfnamefont {H.}~\bibnamefont
  {Choset}}, \ and\ \bibinfo {author} {\bibfnamefont {D.~I.}\ \bibnamefont
  {Goldman}},\ }\href@noop {} {\  (\bibinfo {year} {2019})},\ \Eprint
  {http://arxiv.org/abs/1906.11374} {arXiv:1906.11374 [physics.bio-ph]}
  \BibitemShut {NoStop}%
\bibitem [{\citenamefont {Lauga}\ and\ \citenamefont
  {Powers}(2009)}]{lauga2009hydrodynamics}%
  \BibitemOpen
  \bibfield  {author} {\bibinfo {author} {\bibfnamefont {E.}~\bibnamefont
  {Lauga}}\ and\ \bibinfo {author} {\bibfnamefont {T.~R.}\ \bibnamefont
  {Powers}},\ }\href@noop {} {\bibfield  {journal} {\bibinfo  {journal}
  {Reports on progress in physics}\ }\textbf {\bibinfo {volume} {72}},\
  \bibinfo {pages} {096601} (\bibinfo {year} {2009})}\BibitemShut {NoStop}%
\bibitem [{Note1()}]{Note1}%
  \BibitemOpen
  \bibinfo {note} {Similar principles, often in the form of relationships
  between gait parameters have been previously identified in many different
  categories of locomotion, including bipedal, quadrupedal, hexapodal and
  myriapodal locomotion \cite
  {Heglund1974,Hildebrand1989-rb,Full1991,Chong2023}, flight \cite
  {Sullivan2019}, and aquatic swimming \cite {Gazzola2014,
  Sanchez-Rodriguez2023}}\BibitemShut {NoStop}%
\bibitem [{\citenamefont {Gray}\ and\ \citenamefont
  {Hancock}(1955)}]{gray1955propulsion}%
  \BibitemOpen
  \bibfield  {author} {\bibinfo {author} {\bibfnamefont {J.}~\bibnamefont
  {Gray}}\ and\ \bibinfo {author} {\bibfnamefont {G.}~\bibnamefont {Hancock}},\
  }\href@noop {} {\bibfield  {journal} {\bibinfo  {journal} {Journal of
  Experimental Biology}\ }\textbf {\bibinfo {volume} {32}},\ \bibinfo {pages}
  {802} (\bibinfo {year} {1955})}\BibitemShut {NoStop}%
\bibitem [{\citenamefont {Zhang}\ and\ \citenamefont
  {Goldman}(2014)}]{Zhang2014-fi}%
  \BibitemOpen
  \bibfield  {author} {\bibinfo {author} {\bibfnamefont {T.}~\bibnamefont
  {Zhang}}\ and\ \bibinfo {author} {\bibfnamefont {D.~I.}\ \bibnamefont
  {Goldman}},\ }\href {https://doi.org/10.1063/1.4898629} {\bibfield  {journal}
  {\bibinfo  {journal} {Phys. Fluids}\ }\textbf {\bibinfo {volume} {26}},\
  \bibinfo {pages} {101308} (\bibinfo {year} {2014})}\BibitemShut {NoStop}%
\bibitem [{\citenamefont {Batchelor}(1970)}]{batchelor1970slender}%
  \BibitemOpen
  \bibfield  {author} {\bibinfo {author} {\bibfnamefont {G.~K.}\ \bibnamefont
  {Batchelor}},\ }\href@noop {} {\bibfield  {journal} {\bibinfo  {journal}
  {Journal of Fluid Mechanics}\ }\textbf {\bibinfo {volume} {44}},\ \bibinfo
  {pages} {419} (\bibinfo {year} {1970})}\BibitemShut {NoStop}%
\bibitem [{\citenamefont {Shapere}\ and\ \citenamefont
  {Wilczek}(1989)}]{Shapere1989}%
  \BibitemOpen
  \bibfield  {author} {\bibinfo {author} {\bibfnamefont {A.}~\bibnamefont
  {Shapere}}\ and\ \bibinfo {author} {\bibfnamefont {F.}~\bibnamefont
  {Wilczek}},\ }\href
  {https://www.physics.utoronto.ca/~poppitz/poppitz/PHY1530_files/ShapereLowR.pdf}
  {\bibfield  {journal} {\bibinfo  {journal} {J. Fluid Mech.}\ }\textbf
  {\bibinfo {volume} {198}},\ \bibinfo {pages} {557} (\bibinfo {year}
  {1989})}\BibitemShut {NoStop}%
\bibitem [{\citenamefont {Rieser}\ \emph {et~al.}(2024)\citenamefont {Rieser},
  \citenamefont {Chong}, \citenamefont {Gong}, \citenamefont {Astley},
  \citenamefont {Schiebel}, \citenamefont {Diaz}, \citenamefont {Pierce},
  \citenamefont {Lu}, \citenamefont {Hatton}, \citenamefont {Choset},\ and\
  \citenamefont {Goldman}}]{Rieser2024}%
  \BibitemOpen
  \bibfield  {author} {\bibinfo {author} {\bibfnamefont {J.~M.}\ \bibnamefont
  {Rieser}}, \bibinfo {author} {\bibfnamefont {B.}~\bibnamefont {Chong}},
  \bibinfo {author} {\bibfnamefont {C.}~\bibnamefont {Gong}}, \bibinfo {author}
  {\bibfnamefont {H.~C.}\ \bibnamefont {Astley}}, \bibinfo {author}
  {\bibfnamefont {P.~E.}\ \bibnamefont {Schiebel}}, \bibinfo {author}
  {\bibfnamefont {K.}~\bibnamefont {Diaz}}, \bibinfo {author} {\bibfnamefont
  {C.~J.}\ \bibnamefont {Pierce}}, \bibinfo {author} {\bibfnamefont
  {H.}~\bibnamefont {Lu}}, \bibinfo {author} {\bibfnamefont {R.~L.}\
  \bibnamefont {Hatton}}, \bibinfo {author} {\bibfnamefont {H.}~\bibnamefont
  {Choset}}, \ and\ \bibinfo {author} {\bibfnamefont {D.~I.}\ \bibnamefont
  {Goldman}},\ }\href {http://dx.doi.org/10.1073/pnas.2320517121} {\bibfield
  {journal} {\bibinfo  {journal} {Proc. Natl. Acad. Sci. U. S. A.}\ }\textbf
  {\bibinfo {volume} {121}},\ \bibinfo {pages} {e2320517121} (\bibinfo {year}
  {2024})}\BibitemShut {NoStop}%
\bibitem [{\citenamefont {Schiebel}\ \emph {et~al.}(2020)\citenamefont
  {Schiebel}, \citenamefont {Astley}, \citenamefont {Rieser}, \citenamefont
  {Agarwal}, \citenamefont {Hubicki}, \citenamefont {Hubbard}, \citenamefont
  {Diaz}, \citenamefont {Mendelson}, \citenamefont {Kamrin},\ and\
  \citenamefont {Goldman}}]{Schiebel2020mitigating}%
  \BibitemOpen
  \bibfield  {author} {\bibinfo {author} {\bibfnamefont {P.~E.}\ \bibnamefont
  {Schiebel}}, \bibinfo {author} {\bibfnamefont {H.~C.}\ \bibnamefont
  {Astley}}, \bibinfo {author} {\bibfnamefont {J.~M.}\ \bibnamefont {Rieser}},
  \bibinfo {author} {\bibfnamefont {S.}~\bibnamefont {Agarwal}}, \bibinfo
  {author} {\bibfnamefont {C.}~\bibnamefont {Hubicki}}, \bibinfo {author}
  {\bibfnamefont {A.~M.}\ \bibnamefont {Hubbard}}, \bibinfo {author}
  {\bibfnamefont {K.}~\bibnamefont {Diaz}}, \bibinfo {author} {\bibfnamefont
  {J.~R.}\ \bibnamefont {Mendelson}, \bibfnamefont {Iii}}, \bibinfo {author}
  {\bibfnamefont {K.}~\bibnamefont {Kamrin}}, \ and\ \bibinfo {author}
  {\bibfnamefont {D.~I.}\ \bibnamefont {Goldman}},\ }\href
  {http://dx.doi.org/10.7554/eLife.51412} {\bibfield  {journal} {\bibinfo
  {journal} {Elife}\ }\textbf {\bibinfo {volume} {9}} (\bibinfo {year}
  {2020})}\BibitemShut {NoStop}%
\bibitem [{\citenamefont {McMillen}\ \emph {et~al.}(2008)\citenamefont
  {McMillen}, \citenamefont {Williams},\ and\ \citenamefont
  {Holmes}}]{mcmillen2008nonlinear}%
  \BibitemOpen
  \bibfield  {author} {\bibinfo {author} {\bibfnamefont {T.}~\bibnamefont
  {McMillen}}, \bibinfo {author} {\bibfnamefont {T.}~\bibnamefont {Williams}},
  \ and\ \bibinfo {author} {\bibfnamefont {P.}~\bibnamefont {Holmes}},\
  }\href@noop {} {\bibfield  {journal} {\bibinfo  {journal} {PLoS computational
  biology}\ }\textbf {\bibinfo {volume} {4}},\ \bibinfo {pages} {e1000157}
  (\bibinfo {year} {2008})}\BibitemShut {NoStop}%
\bibitem [{\citenamefont {Ding}\ \emph {et~al.}(2013)\citenamefont {Ding},
  \citenamefont {Sharpe}, \citenamefont {Wiesenfeld},\ and\ \citenamefont
  {Goldman}}]{ding2013emergence}%
  \BibitemOpen
  \bibfield  {author} {\bibinfo {author} {\bibfnamefont {Y.}~\bibnamefont
  {Ding}}, \bibinfo {author} {\bibfnamefont {S.~S.}\ \bibnamefont {Sharpe}},
  \bibinfo {author} {\bibfnamefont {K.}~\bibnamefont {Wiesenfeld}}, \ and\
  \bibinfo {author} {\bibfnamefont {D.~I.}\ \bibnamefont {Goldman}},\
  }\href@noop {} {\bibfield  {journal} {\bibinfo  {journal} {Proceedings of the
  National Academy of Sciences}\ }\textbf {\bibinfo {volume} {110}},\ \bibinfo
  {pages} {10123} (\bibinfo {year} {2013})}\BibitemShut {NoStop}%
\bibitem [{\citenamefont {Pierce}\ \emph {et~al.}(2025)\citenamefont {Pierce},
  \citenamefont {Ding}, \citenamefont {Peng}, \citenamefont {Lu}, \citenamefont
  {Chong}, \citenamefont {Lu},\ and\ \citenamefont
  {Goldman}}]{pierce2025neuromechanical}%
  \BibitemOpen
  \bibfield  {author} {\bibinfo {author} {\bibfnamefont {C.~J.}\ \bibnamefont
  {Pierce}}, \bibinfo {author} {\bibfnamefont {Y.}~\bibnamefont {Ding}},
  \bibinfo {author} {\bibfnamefont {L.}~\bibnamefont {Peng}}, \bibinfo {author}
  {\bibfnamefont {X.}~\bibnamefont {Lu}}, \bibinfo {author} {\bibfnamefont
  {B.}~\bibnamefont {Chong}}, \bibinfo {author} {\bibfnamefont
  {H.}~\bibnamefont {Lu}}, \ and\ \bibinfo {author} {\bibfnamefont {D.~I.}\
  \bibnamefont {Goldman}},\ }\href@noop {} {\bibfield  {journal} {\bibinfo
  {journal} {PRX Life}\ }\textbf {\bibinfo {volume} {3}},\ \bibinfo {pages}
  {023001} (\bibinfo {year} {2025})}\BibitemShut {NoStop}%
\bibitem [{\citenamefont {Butler}\ \emph {et~al.}(2015)\citenamefont {Butler},
  \citenamefont {Branicky}, \citenamefont {Yemini}, \citenamefont {Liewald},
  \citenamefont {Gottschalk}, \citenamefont {Kerr}, \citenamefont
  {Chklovskii},\ and\ \citenamefont {Schafer}}]{butler2015consistent}%
  \BibitemOpen
  \bibfield  {author} {\bibinfo {author} {\bibfnamefont {V.~J.}\ \bibnamefont
  {Butler}}, \bibinfo {author} {\bibfnamefont {R.}~\bibnamefont {Branicky}},
  \bibinfo {author} {\bibfnamefont {E.}~\bibnamefont {Yemini}}, \bibinfo
  {author} {\bibfnamefont {J.~F.}\ \bibnamefont {Liewald}}, \bibinfo {author}
  {\bibfnamefont {A.}~\bibnamefont {Gottschalk}}, \bibinfo {author}
  {\bibfnamefont {R.~A.}\ \bibnamefont {Kerr}}, \bibinfo {author}
  {\bibfnamefont {D.~B.}\ \bibnamefont {Chklovskii}}, \ and\ \bibinfo {author}
  {\bibfnamefont {W.~R.}\ \bibnamefont {Schafer}},\ }\href@noop {} {\bibfield
  {journal} {\bibinfo  {journal} {Journal of the Royal Society Interface}\
  }\textbf {\bibinfo {volume} {12}},\ \bibinfo {pages} {20140963} (\bibinfo
  {year} {2015})}\BibitemShut {NoStop}%
\bibitem [{\citenamefont {Fang-Yen}\ \emph {et~al.}(2010)\citenamefont
  {Fang-Yen}, \citenamefont {Wyart}, \citenamefont {Xie}, \citenamefont
  {Kawai}, \citenamefont {Kodger}, \citenamefont {Chen}, \citenamefont {Wen},\
  and\ \citenamefont {Samuel}}]{fang2010biomechanical}%
  \BibitemOpen
  \bibfield  {author} {\bibinfo {author} {\bibfnamefont {C.}~\bibnamefont
  {Fang-Yen}}, \bibinfo {author} {\bibfnamefont {M.}~\bibnamefont {Wyart}},
  \bibinfo {author} {\bibfnamefont {J.}~\bibnamefont {Xie}}, \bibinfo {author}
  {\bibfnamefont {R.}~\bibnamefont {Kawai}}, \bibinfo {author} {\bibfnamefont
  {T.}~\bibnamefont {Kodger}}, \bibinfo {author} {\bibfnamefont
  {S.}~\bibnamefont {Chen}}, \bibinfo {author} {\bibfnamefont {Q.}~\bibnamefont
  {Wen}}, \ and\ \bibinfo {author} {\bibfnamefont {A.~D.}\ \bibnamefont
  {Samuel}},\ }\href@noop {} {\bibfield  {journal} {\bibinfo  {journal}
  {Proceedings of the National Academy of Sciences}\ }\textbf {\bibinfo
  {volume} {107}},\ \bibinfo {pages} {20323} (\bibinfo {year}
  {2010})}\BibitemShut {NoStop}%
\bibitem [{\citenamefont {F{\'e}lix}\ and\ \citenamefont
  {Braendle}(2010)}]{felix2010natural}%
  \BibitemOpen
  \bibfield  {author} {\bibinfo {author} {\bibfnamefont {M.-A.}\ \bibnamefont
  {F{\'e}lix}}\ and\ \bibinfo {author} {\bibfnamefont {C.}~\bibnamefont
  {Braendle}},\ }\href@noop {} {\bibfield  {journal} {\bibinfo  {journal}
  {Current biology}\ }\textbf {\bibinfo {volume} {20}},\ \bibinfo {pages}
  {R965} (\bibinfo {year} {2010})}\BibitemShut {NoStop}%
\bibitem [{\citenamefont {Juarez}\ \emph {et~al.}(2010)\citenamefont {Juarez},
  \citenamefont {Lu}, \citenamefont {Sznitman},\ and\ \citenamefont
  {Arratia}}]{juarez2010motility}%
  \BibitemOpen
  \bibfield  {author} {\bibinfo {author} {\bibfnamefont {G.}~\bibnamefont
  {Juarez}}, \bibinfo {author} {\bibfnamefont {K.}~\bibnamefont {Lu}}, \bibinfo
  {author} {\bibfnamefont {J.}~\bibnamefont {Sznitman}}, \ and\ \bibinfo
  {author} {\bibfnamefont {P.~E.}\ \bibnamefont {Arratia}},\ }\href@noop {}
  {\bibfield  {journal} {\bibinfo  {journal} {Europhysics Letters}\ }\textbf
  {\bibinfo {volume} {92}},\ \bibinfo {pages} {44002} (\bibinfo {year}
  {2010})}\BibitemShut {NoStop}%
\bibitem [{\citenamefont {Shen}\ and\ \citenamefont
  {Arratia}(2011)}]{shen2011undulatory}%
  \BibitemOpen
  \bibfield  {author} {\bibinfo {author} {\bibfnamefont {X.}~\bibnamefont
  {Shen}}\ and\ \bibinfo {author} {\bibfnamefont {P.~E.}\ \bibnamefont
  {Arratia}},\ }\href@noop {} {\bibfield  {journal} {\bibinfo  {journal}
  {Physical review letters}\ }\textbf {\bibinfo {volume} {106}},\ \bibinfo
  {pages} {208101} (\bibinfo {year} {2011})}\BibitemShut {NoStop}%
\bibitem [{\citenamefont {Majmudar}\ \emph {et~al.}(2012)\citenamefont
  {Majmudar}, \citenamefont {Keaveny}, \citenamefont {Zhang},\ and\
  \citenamefont {Shelley}}]{majmudar2012experiments}%
  \BibitemOpen
  \bibfield  {author} {\bibinfo {author} {\bibfnamefont {T.}~\bibnamefont
  {Majmudar}}, \bibinfo {author} {\bibfnamefont {E.~E.}\ \bibnamefont
  {Keaveny}}, \bibinfo {author} {\bibfnamefont {J.}~\bibnamefont {Zhang}}, \
  and\ \bibinfo {author} {\bibfnamefont {M.~J.}\ \bibnamefont {Shelley}},\
  }\href@noop {} {\bibfield  {journal} {\bibinfo  {journal} {Journal of the
  Royal Society Interface}\ }\textbf {\bibinfo {volume} {9}},\ \bibinfo {pages}
  {1809} (\bibinfo {year} {2012})}\BibitemShut {NoStop}%
\bibitem [{\citenamefont {Wang}\ \emph {et~al.}(2023)\citenamefont {Wang},
  \citenamefont {Pierce}, \citenamefont {Kojouharov}, \citenamefont {Chong},
  \citenamefont {Diaz}, \citenamefont {Lu},\ and\ \citenamefont
  {Goldman}}]{Wang2023mechanical}%
  \BibitemOpen
  \bibfield  {author} {\bibinfo {author} {\bibfnamefont {T.}~\bibnamefont
  {Wang}}, \bibinfo {author} {\bibfnamefont {C.}~\bibnamefont {Pierce}},
  \bibinfo {author} {\bibfnamefont {V.}~\bibnamefont {Kojouharov}}, \bibinfo
  {author} {\bibfnamefont {B.}~\bibnamefont {Chong}}, \bibinfo {author}
  {\bibfnamefont {K.}~\bibnamefont {Diaz}}, \bibinfo {author} {\bibfnamefont
  {H.}~\bibnamefont {Lu}}, \ and\ \bibinfo {author} {\bibfnamefont {D.~I.}\
  \bibnamefont {Goldman}},\ }\href@noop {} {\bibfield  {journal} {\bibinfo
  {journal} {Sci Robot}\ }\textbf {\bibinfo {volume} {8}},\ \bibinfo {pages}
  {eadi2243} (\bibinfo {year} {2023})}\BibitemShut {NoStop}%
\bibitem [{Note2()}]{Note2}%
  \BibitemOpen
  \bibinfo {note} {To avoid confusion, we use the following notational
  convention throughout the manuscript. For dimensionless quantities, we use
  the angular frequency and wavenumber $\omega $ and $k$. When presenting
  experimental data, we use the corresponding linear frequency and inverse
  wavelength in SI units, denoted $f$ and $1/\lambda $,
  respectively.}\BibitemShut {Stop}%
\bibitem [{\citenamefont {Guo}\ and\ \citenamefont
  {Mahadevan}(2008)}]{Guo2008}%
  \BibitemOpen
  \bibfield  {author} {\bibinfo {author} {\bibfnamefont {Z.~V.}\ \bibnamefont
  {Guo}}\ and\ \bibinfo {author} {\bibfnamefont {L.}~\bibnamefont
  {Mahadevan}},\ }\href {http://dx.doi.org/10.1073/pnas.0705442105} {\bibfield
  {journal} {\bibinfo  {journal} {Proc. Natl. Acad. Sci. U. S. A.}\ }\textbf
  {\bibinfo {volume} {105}},\ \bibinfo {pages} {3179} (\bibinfo {year}
  {2008})}\BibitemShut {NoStop}%
\bibitem [{Note3()}]{Note3}%
  \BibitemOpen
  \bibinfo {note} {In the small amplitude limit used in this manuscript, the
  normal fluid drag is approximated by $C \protect \dot {y}$ and the
  $y$-component of the tangential fluid drag is negligible.}\BibitemShut
  {Stop}%
\bibitem [{Note4()}]{Note4}%
  \BibitemOpen
  \bibinfo {note} {The full equation of motion and both limiting cases produce
  the same wave stability criterion.}\BibitemShut {Stop}%
\bibitem [{\citenamefont {McMillen}\ and\ \citenamefont
  {Holmes}(2006)}]{McMillen2006elastic}%
  \BibitemOpen
  \bibfield  {author} {\bibinfo {author} {\bibfnamefont {T.}~\bibnamefont
  {McMillen}}\ and\ \bibinfo {author} {\bibfnamefont {P.}~\bibnamefont
  {Holmes}},\ }\href {http://dx.doi.org/10.1007/s00285-006-0036-8} {\bibfield
  {journal} {\bibinfo  {journal} {J. Math. Biol.}\ }\textbf {\bibinfo {volume}
  {53}},\ \bibinfo {pages} {843} (\bibinfo {year} {2006})}\BibitemShut
  {NoStop}%
\bibitem [{\citenamefont {Lighthill}(1960)}]{Lighthill1960note}%
  \BibitemOpen
  \bibfield  {author} {\bibinfo {author} {\bibfnamefont {M.~J.}\ \bibnamefont
  {Lighthill}},\ }\href {http://dx.doi.org/10.1017/S0022112060001110}
  {\bibfield  {journal} {\bibinfo  {journal} {J. Fluid Mech.}\ }\textbf
  {\bibinfo {volume} {9}},\ \bibinfo {pages} {305} (\bibinfo {year}
  {1960})}\BibitemShut {NoStop}%
\bibitem [{\citenamefont {Hess}\ and\ \citenamefont
  {Videler}(1984)}]{Hess1984fast}%
  \BibitemOpen
  \bibfield  {author} {\bibinfo {author} {\bibfnamefont {F.}~\bibnamefont
  {Hess}}\ and\ \bibinfo {author} {\bibfnamefont {J.~J.}\ \bibnamefont
  {Videler}},\ }\href {https://dx.doi.org/10.1242/jeb.109.1.229} {\bibfield
  {journal} {\bibinfo  {journal} {J. Exp. Biol.}\ }\textbf {\bibinfo {volume}
  {109}},\ \bibinfo {pages} {229} (\bibinfo {year} {1984})}\BibitemShut
  {NoStop}%
\bibitem [{\citenamefont {Cheng}\ \emph {et~al.}(1998)\citenamefont {Cheng},
  \citenamefont {Pedley},\ and\ \citenamefont
  {Altringham}}]{Cheng1998analysis}%
  \BibitemOpen
  \bibfield  {author} {\bibinfo {author} {\bibfnamefont {J.~ .}\ \bibnamefont
  {Cheng}}, \bibinfo {author} {\bibfnamefont {T.~J.}\ \bibnamefont {Pedley}}, \
  and\ \bibinfo {author} {\bibfnamefont {J.~D.}\ \bibnamefont {Altringham}},\
  }\href {http://dx.doi.org/10.1098/rstb.1998.0262} {\bibfield  {journal}
  {\bibinfo  {journal} {Philos. Trans. R. Soc. Lond. B Biol. Sci.}\ }\textbf
  {\bibinfo {volume} {353}},\ \bibinfo {pages} {981} (\bibinfo {year}
  {1998})}\BibitemShut {NoStop}%
\bibitem [{\citenamefont {van Weerden}\ \emph {et~al.}(2014)\citenamefont {van
  Weerden}, \citenamefont {Reid},\ and\ \citenamefont
  {Hemelrijk}}]{Van_Weerden2014}%
  \BibitemOpen
  \bibfield  {author} {\bibinfo {author} {\bibfnamefont {J.~F.}\ \bibnamefont
  {van Weerden}}, \bibinfo {author} {\bibfnamefont {D.~A.~P.}\ \bibnamefont
  {Reid}}, \ and\ \bibinfo {author} {\bibfnamefont {C.~K.}\ \bibnamefont
  {Hemelrijk}},\ }\href {https://onlinelibrary.wiley.com/doi/10.1111/faf.12022}
  {\bibfield  {journal} {\bibinfo  {journal} {Fish Fish}\ }\textbf {\bibinfo
  {volume} {15}},\ \bibinfo {pages} {397} (\bibinfo {year} {2014})}\BibitemShut
  {NoStop}%
\bibitem [{\citenamefont {Velho~Rodrigues}\ \emph {et~al.}(2021)\citenamefont
  {Velho~Rodrigues}, \citenamefont {Lisicki},\ and\ \citenamefont
  {Lauga}}]{Velho_Rodrigues2021}%
  \BibitemOpen
  \bibfield  {author} {\bibinfo {author} {\bibfnamefont {M.~F.}\ \bibnamefont
  {Velho~Rodrigues}}, \bibinfo {author} {\bibfnamefont {M.}~\bibnamefont
  {Lisicki}}, \ and\ \bibinfo {author} {\bibfnamefont {E.}~\bibnamefont
  {Lauga}},\ }\href {http://dx.doi.org/10.1371/journal.pone.0252291} {\bibfield
   {journal} {\bibinfo  {journal} {PLoS One}\ }\textbf {\bibinfo {volume}
  {16}},\ \bibinfo {pages} {e0252291} (\bibinfo {year} {2021})}\BibitemShut
  {NoStop}%
\bibitem [{\citenamefont {Gray}\ and\ \citenamefont
  {Lissmann}(1964)}]{Gray1964locomotion}%
  \BibitemOpen
  \bibfield  {author} {\bibinfo {author} {\bibfnamefont {J.}~\bibnamefont
  {Gray}}\ and\ \bibinfo {author} {\bibfnamefont {H.~W.}\ \bibnamefont
  {Lissmann}},\ }\href@noop {} {\bibfield  {journal} {\bibinfo  {journal} {J.
  Exp. Biol.}\ }\textbf {\bibinfo {volume} {41}},\ \bibinfo {pages} {135}
  (\bibinfo {year} {1964})}\BibitemShut {NoStop}%
\bibitem [{\citenamefont {Dorgan}\ \emph {et~al.}(2013)\citenamefont {Dorgan},
  \citenamefont {Law},\ and\ \citenamefont {Rouse}}]{Dorgan2013}%
  \BibitemOpen
  \bibfield  {author} {\bibinfo {author} {\bibfnamefont {K.~M.}\ \bibnamefont
  {Dorgan}}, \bibinfo {author} {\bibfnamefont {C.~J.}\ \bibnamefont {Law}}, \
  and\ \bibinfo {author} {\bibfnamefont {G.~W.}\ \bibnamefont {Rouse}},\ }\href
  {http://dx.doi.org/10.1098/rspb.2012.2948} {\bibfield  {journal} {\bibinfo
  {journal} {Proc. Biol. Sci.}\ }\textbf {\bibinfo {volume} {280}},\ \bibinfo
  {pages} {20122948} (\bibinfo {year} {2013})}\BibitemShut {NoStop}%
\bibitem [{\citenamefont {Maes}\ \emph {et~al.}(2012)\citenamefont {Maes},
  \citenamefont {Verlooy}, \citenamefont {Buenafe}, \citenamefont {de~Witte},
  \citenamefont {Esguerra},\ and\ \citenamefont {Crawford}}]{Maes2012}%
  \BibitemOpen
  \bibfield  {author} {\bibinfo {author} {\bibfnamefont {J.}~\bibnamefont
  {Maes}}, \bibinfo {author} {\bibfnamefont {L.}~\bibnamefont {Verlooy}},
  \bibinfo {author} {\bibfnamefont {O.~E.}\ \bibnamefont {Buenafe}}, \bibinfo
  {author} {\bibfnamefont {P.~A.~M.}\ \bibnamefont {de~Witte}}, \bibinfo
  {author} {\bibfnamefont {C.~V.}\ \bibnamefont {Esguerra}}, \ and\ \bibinfo
  {author} {\bibfnamefont {A.~D.}\ \bibnamefont {Crawford}},\ }\href
  {http://dx.doi.org/10.1371/journal.pone.0043850} {\bibfield  {journal}
  {\bibinfo  {journal} {PLoS One}\ }\textbf {\bibinfo {volume} {7}},\ \bibinfo
  {pages} {e43850} (\bibinfo {year} {2012})}\BibitemShut {NoStop}%
\bibitem [{\citenamefont {Fischbach}\ \emph {et~al.}(2023)\citenamefont
  {Fischbach}, \citenamefont {Finke}, \citenamefont {Moritz}, \citenamefont
  {Polte},\ and\ \citenamefont {Thieme}}]{Fischbach2023}%
  \BibitemOpen
  \bibfield  {author} {\bibinfo {author} {\bibfnamefont {V.}~\bibnamefont
  {Fischbach}}, \bibinfo {author} {\bibfnamefont {A.}~\bibnamefont {Finke}},
  \bibinfo {author} {\bibfnamefont {T.}~\bibnamefont {Moritz}}, \bibinfo
  {author} {\bibfnamefont {P.}~\bibnamefont {Polte}}, \ and\ \bibinfo {author}
  {\bibfnamefont {P.}~\bibnamefont {Thieme}},\ }\href
  {https://aslopubs.onlinelibrary.wiley.com/doi/10.1002/lom3.10551} {\bibfield
  {journal} {\bibinfo  {journal} {Limnol. Oceanogr. Methods}\ }\textbf
  {\bibinfo {volume} {21}},\ \bibinfo {pages} {357} (\bibinfo {year}
  {2023})}\BibitemShut {NoStop}%
\bibitem [{\citenamefont {Rikmenspoel}(1978)}]{Rikmenspoel1978}%
  \BibitemOpen
  \bibfield  {author} {\bibinfo {author} {\bibfnamefont {R.}~\bibnamefont
  {Rikmenspoel}},\ }\href {http://dx.doi.org/10.1016/S0006-3495(78)85442-3}
  {\bibfield  {journal} {\bibinfo  {journal} {Biophys. J.}\ }\textbf {\bibinfo
  {volume} {23}},\ \bibinfo {pages} {177} (\bibinfo {year} {1978})}\BibitemShut
  {NoStop}%
\bibitem [{\citenamefont {Panton}(2024)}]{PantonHydro}%
  \BibitemOpen
  \bibfield  {author} {\bibinfo {author} {\bibfnamefont {R.~L.}\ \bibnamefont
  {Panton}},\ }\href@noop {} {\emph {\bibinfo {title} {Incompressible Flow, 5th
  Edition}}}\ (\bibinfo  {publisher} {Wiley},\ \bibinfo {year}
  {2024})\BibitemShut {NoStop}%
\bibitem [{\citenamefont {Ji}\ \emph {et~al.}(2021)\citenamefont {Ji},
  \citenamefont {Fouad}, \citenamefont {Teng}, \citenamefont {Liu},
  \citenamefont {Alvarez-Illera}, \citenamefont {Yao}, \citenamefont {Li},\
  and\ \citenamefont {Fang-Yen}}]{Ji2021}%
  \BibitemOpen
  \bibfield  {author} {\bibinfo {author} {\bibfnamefont {H.}~\bibnamefont
  {Ji}}, \bibinfo {author} {\bibfnamefont {A.~D.}\ \bibnamefont {Fouad}},
  \bibinfo {author} {\bibfnamefont {S.}~\bibnamefont {Teng}}, \bibinfo {author}
  {\bibfnamefont {A.}~\bibnamefont {Liu}}, \bibinfo {author} {\bibfnamefont
  {P.}~\bibnamefont {Alvarez-Illera}}, \bibinfo {author} {\bibfnamefont
  {B.}~\bibnamefont {Yao}}, \bibinfo {author} {\bibfnamefont {Z.}~\bibnamefont
  {Li}}, \ and\ \bibinfo {author} {\bibfnamefont {C.}~\bibnamefont
  {Fang-Yen}},\ }\href {http://dx.doi.org/10.7554/eLife.69905} {\bibfield
  {journal} {\bibinfo  {journal} {Elife}\ }\textbf {\bibinfo {volume} {10}}
  (\bibinfo {year} {2021})}\BibitemShut {NoStop}%
\bibitem [{\citenamefont {Johnson}\ \emph {et~al.}(2021)\citenamefont
  {Johnson}, \citenamefont {Lewis},\ and\ \citenamefont
  {Guy}}]{johnson2021neuromechanical}%
  \BibitemOpen
  \bibfield  {author} {\bibinfo {author} {\bibfnamefont {C.~L.}\ \bibnamefont
  {Johnson}}, \bibinfo {author} {\bibfnamefont {T.~J.}\ \bibnamefont {Lewis}},
  \ and\ \bibinfo {author} {\bibfnamefont {R.}~\bibnamefont {Guy}},\
  }\href@noop {} {\bibfield  {journal} {\bibinfo  {journal} {SIAM Journal on
  Applied Dynamical Systems}\ }\textbf {\bibinfo {volume} {20}},\ \bibinfo
  {pages} {1022} (\bibinfo {year} {2021})}\BibitemShut {NoStop}%
\bibitem [{\citenamefont {Heglund}\ \emph {et~al.}(1974)\citenamefont
  {Heglund}, \citenamefont {Taylor},\ and\ \citenamefont
  {McMahon}}]{Heglund1974}%
  \BibitemOpen
  \bibfield  {author} {\bibinfo {author} {\bibfnamefont {N.~C.}\ \bibnamefont
  {Heglund}}, \bibinfo {author} {\bibfnamefont {C.~R.}\ \bibnamefont {Taylor}},
  \ and\ \bibinfo {author} {\bibfnamefont {T.~A.}\ \bibnamefont {McMahon}},\
  }\href {http://dx.doi.org/10.1126/science.186.4169.1112} {\bibfield
  {journal} {\bibinfo  {journal} {Science}\ }\textbf {\bibinfo {volume}
  {186}},\ \bibinfo {pages} {1112} (\bibinfo {year} {1974})}\BibitemShut
  {NoStop}%
\bibitem [{\citenamefont {Hildebrand}(1989)}]{Hildebrand1989-rb}%
  \BibitemOpen
  \bibfield  {author} {\bibinfo {author} {\bibfnamefont {M.}~\bibnamefont
  {Hildebrand}},\ }\href@noop {} {\bibfield  {journal} {\bibinfo  {journal}
  {Bioscience}\ }\textbf {\bibinfo {volume} {39}},\ \bibinfo {pages} {766}
  (\bibinfo {year} {1989})}\BibitemShut {NoStop}%
\bibitem [{\citenamefont {Full}\ and\ \citenamefont {Tu}(1991)}]{Full1991}%
  \BibitemOpen
  \bibfield  {author} {\bibinfo {author} {\bibfnamefont {R.~J.}\ \bibnamefont
  {Full}}\ and\ \bibinfo {author} {\bibfnamefont {M.~S.}\ \bibnamefont {Tu}},\
  }\href {http://dx.doi.org/10.1242/jeb.156.1.215} {\bibfield  {journal}
  {\bibinfo  {journal} {J. Exp. Biol.}\ }\textbf {\bibinfo {volume} {156}},\
  \bibinfo {pages} {215} (\bibinfo {year} {1991})}\BibitemShut {NoStop}%
\bibitem [{\citenamefont {Chong}\ \emph
  {et~al.}(2023{\natexlab{b}})\citenamefont {Chong}, \citenamefont {He},
  \citenamefont {Soto}, \citenamefont {Wang}, \citenamefont {Irvine},
  \citenamefont {Blekherman},\ and\ \citenamefont {Goldman}}]{Chong2023}%
  \BibitemOpen
  \bibfield  {author} {\bibinfo {author} {\bibfnamefont {B.}~\bibnamefont
  {Chong}}, \bibinfo {author} {\bibfnamefont {J.}~\bibnamefont {He}}, \bibinfo
  {author} {\bibfnamefont {D.}~\bibnamefont {Soto}}, \bibinfo {author}
  {\bibfnamefont {T.}~\bibnamefont {Wang}}, \bibinfo {author} {\bibfnamefont
  {D.}~\bibnamefont {Irvine}}, \bibinfo {author} {\bibfnamefont
  {G.}~\bibnamefont {Blekherman}}, \ and\ \bibinfo {author} {\bibfnamefont
  {D.~I.}\ \bibnamefont {Goldman}},\ }\href@noop {} {\bibfield  {journal}
  {\bibinfo  {journal} {Science}\ }\textbf {\bibinfo {volume} {380}},\ \bibinfo
  {pages} {509} (\bibinfo {year} {2023}{\natexlab{b}})}\BibitemShut {NoStop}%
\bibitem [{\citenamefont {Sullivan}\ \emph {et~al.}(2019)\citenamefont
  {Sullivan}, \citenamefont {Meyers},\ and\ \citenamefont
  {Arzt}}]{Sullivan2019}%
  \BibitemOpen
  \bibfield  {author} {\bibinfo {author} {\bibfnamefont {T.~N.}\ \bibnamefont
  {Sullivan}}, \bibinfo {author} {\bibfnamefont {M.~A.}\ \bibnamefont
  {Meyers}}, \ and\ \bibinfo {author} {\bibfnamefont {E.}~\bibnamefont
  {Arzt}},\ }\href {http://dx.doi.org/10.1126/sciadv.aat4269} {\bibfield
  {journal} {\bibinfo  {journal} {Sci Adv}\ }\textbf {\bibinfo {volume} {5}},\
  \bibinfo {pages} {eaat4269} (\bibinfo {year} {2019})}\BibitemShut {NoStop}%
\bibitem [{\citenamefont {Gazzola}\ \emph {et~al.}(2014)\citenamefont
  {Gazzola}, \citenamefont {Argentina},\ and\ \citenamefont
  {Mahadevan}}]{Gazzola2014}%
  \BibitemOpen
  \bibfield  {author} {\bibinfo {author} {\bibfnamefont {M.}~\bibnamefont
  {Gazzola}}, \bibinfo {author} {\bibfnamefont {M.}~\bibnamefont {Argentina}},
  \ and\ \bibinfo {author} {\bibfnamefont {L.}~\bibnamefont {Mahadevan}},\
  }\href {https://www.nature.com/articles/nphys3078} {\bibfield  {journal}
  {\bibinfo  {journal} {Nat. Phys.}\ }\textbf {\bibinfo {volume} {10}},\
  \bibinfo {pages} {758} (\bibinfo {year} {2014})}\BibitemShut {NoStop}%
\bibitem [{\citenamefont {S{\'a}nchez-Rodr{\'\i}guez}\ \emph
  {et~al.}(2023)\citenamefont {S{\'a}nchez-Rodr{\'\i}guez}, \citenamefont
  {Raufaste},\ and\ \citenamefont {Argentina}}]{Sanchez-Rodriguez2023}%
  \BibitemOpen
  \bibfield  {author} {\bibinfo {author} {\bibfnamefont {J.}~\bibnamefont
  {S{\'a}nchez-Rodr{\'\i}guez}}, \bibinfo {author} {\bibfnamefont
  {C.}~\bibnamefont {Raufaste}}, \ and\ \bibinfo {author} {\bibfnamefont
  {M.}~\bibnamefont {Argentina}},\ }\href
  {http://dx.doi.org/10.1038/s41467-023-41368-6} {\bibfield  {journal}
  {\bibinfo  {journal} {Nat. Commun.}\ }\textbf {\bibinfo {volume} {14}},\
  \bibinfo {pages} {5569} (\bibinfo {year} {2023})}\BibitemShut {NoStop}%
\end{thebibliography}
\end{document}